# Heteroclinic-structure transition of the pure quartic modulation instability

Xiankun Yao,[1,2,3,*] Chong Liu,[1,2,3,†] Zhan-Ying Yang,[1,2,3,‡] and Wen-Li Yang[1,2,3,4]

[1]*School of Physics, Northwest University, 710127 Xi'an, People's Republic of China*
[2]*Shaanxi Key Laboratory for Theoretical Physics Frontiers, 710127 Xi'an, People's Republic of China*
[3]*Peng Huanwu Center for Fundamental Theory, 710127 Xi'an, People's Republic of China*
[4]*Institute of Modern Physics, Northwest University, 710127 Xi'an, People's Republic of China*

We show that, in the pure-quartic systems, modulation instability (MI) undergoes heteroclinic-structure transitions (HSTs) at two critical frequencies of $\omega_{c1}$ and $\omega_{c2}$ ($\omega_{c2} > \omega_{c1}$), which indicates that there are significant changes of the spatiotemporal behavior in the system. The complicated heteroclinic structure of instability obtained by the mode truncation method reveals all possible dynamic trajectories of nonlinear waves, which allows us to discover the various types of Fermi-Pasta-Ulam (FPU) recurrences and Akhmediev breathers (ABs). When the modulational frequency satisfies $\omega < \omega_{c2}$, the heteroclinic structure encompasses two separatrixes corresponding to the ABs and the nonlinear wave with a modulated final state, which individually separate FPU recurrences into three different regions. Remarkably, crossing critical frequency $\omega_{c1}$, both the staggered FPU recurrences and ABs essentially switch their patterns. These HST behaviors will give vitality to the study of MI.

DOI: 10.1103/PhysRevResearch.4.013246

## I. INTRODUCTION

Modulation instability (MI), the key mechanism for the spontaneous nonlinear evolution of perturbed homogeneous states into complex patterns, is responsible for the formation of many complex patterns such as Fermi-Pasta-Ulam (FPU) recurrences [1–4], Akhmediev breathers (ABs) [5–8], Kuznetsov-Ma breathers [9,10], superregular breathers [11–13], and rogue waves [14–18]. When taking the weakly modulated monochromatic continuous wave (CW) as an initial condition, the long-term evolution of MI has a complex spatiotemporal dynamics that exhibits fierce power exchange between the CW pump and spectral sidebands via a cascade four-wave mixing process [19–21]. This frequency-conversion process of MI can be efficiently characterized by the Hamiltonian contours on the phase plane that refer to the heteroclinic structure by using the multiwave truncation method [22,23]. Generally, there are separatrixes on the heteroclinic structure corresponding to ABs which divide the phase plane into different FPU recurrent regimes [3,24]. Especially, in dispersion-oscillating and high-birefringence fiber systems, the topology of the heteroclinic structure changes radically when switching the sideband frequency across the upper edge of gain bandwidth [25–28], i.e., the heteroclinic-structure transition (HST) happens. Currently, research on HST is limited to the simple frequency-conversion characteristics, while what essential effects the HST can bring to the time-domain patterns remains entirely unexplored. Here, we will show that the HST can greatly enrich the diversity of nonlinear local waves.

Recently, the biharmonic nonlinear Schrödinger equation (NSE), a generalized NSE with the combined effects of quartic dispersion and a Kerr nonlinearity, has attracted intense attention due to its breakthrough applications in pure-quartic solitons [29–33]. Here, we choose the biharmonic NSE as a physical model to study MI for two reasons. First, the equation has been widely implemented in experiments. At present, many optical platforms, such as silicon photonic crystal waveguides [29] and silica photonic crystal fibers [30], are proved to be capable of realizing the biharmonic NSE. Second, benefitting from its fourth-order dispersion effect, the biharmonic NSE can produce HST at the same time without higher-order MI occurring, thus ensuring the applicability of the mode truncation method.

In this paper, we investigate the evolution characteristics of initially modulated plane waves in the biharmonic NSE. Through the phase-space structure, we show that MI undergoes HST at two critical frequencies of $\omega_{c1}$ and $\omega_{c2}$ ($\omega_{c2} > \omega_{c1}$). When modulational frequency is below $\omega_{c2}$, the modulated plane waves can evolve into various nonlinear local waves, including FPU recurrences and ABs and the nonlinear wave with the modulated final state. Both the staggered-FPU recurrences and ABs can switch their patterns across the minor critical frequency $\omega_{c1}$.

## II. PHYSICAL MODEL AND LINEAR STABILITY ANALYSIS

The optical-wave evolution is governed by the following dimensionless biharmonic NSE:

---

*yaoxk@nwu.edu.cn
†chongliu@nwu.edu.cn
‡zyyang@nwu.edu.cn

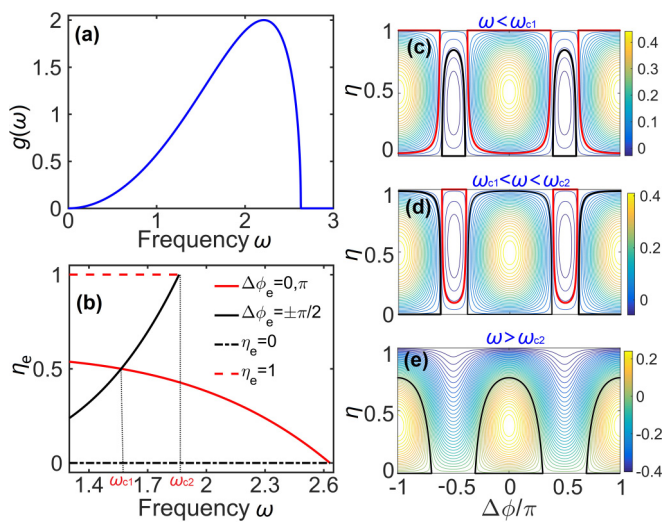

FIG. 1. (a) Modulation instability (MI) gain spectrum profile for Eq. (1). (b) Stationary sideband fraction $\eta_e$ of stable (solid lines) and unstable (dash and dash-dotted lines) branches vs $\omega$. Hamiltonian contours on the phase plane ($\Delta\phi$, $\eta$) for modulational frequencies (c) $\omega = 1.5$ below $\omega_{c1}$, (d) $\omega = 1.6$ inside [$\omega_{c1}$, $\omega_{c2}$] range, and (e) $\omega = 2$ beyond $\omega_{c2}$, respectively. The black and red curves in (c)–(e) are the heteroclinic separatrixes with $\eta_e = 1$ and 0, respectively. Here, $P = 1$.

$$iA_z + \frac{\beta_4}{24}\frac{\partial^4 A}{\partial t^4} + |A|^2 A = 0, \quad (1)$$

where $z$ and $t$ are normalized distance and time, respectively. Here, $\beta_4$ is the four-order dispersion coefficient. This equation governs the nonlinear systems only possessing quartic dispersion and the Kerr effect. In this paper, only the anomalous-dispersion case is discussed, and $\beta_4$ is valued as $-1$ in all calculations. Equation (1) can be realized in a recirculating fiber loop like the one reported in Ref. [34]. In the fiber loop, the linear losses are compensated by Raman amplification. The pure quartic dispersion can be achieved by compensating the quadratic and cubic dispersion through inserting a spatial light modulator in the fiber loop.

Equation (1) has a stationary solution $A_0(z, t) = \sqrt{P} \exp(iPz)$ with power $P = |A_0|^2$. The MI is analyzed by the nonlinear evolution of the perturbed stationary solution $A = A_0[1 + a(z, t)]$, $a$ being a perturbation at modulational frequency $\omega$ and wave number $K$ of the form $a = \{\varepsilon_1 \exp[i(Kz-\omega t)] + \varepsilon_2 \exp[-i(Kz-\omega t)]\}$. Through linear stability analysis, we obtain the dispersion relation $K = \frac{|\beta_4|\omega^2}{24}(\omega^4 + 48P/\beta_4)^{1/2}$. The instability gain can be obtained by $g(\omega) = 2\text{Im}(K)$. Figure 1(a) displays the MI gain spectrum (only the $\omega > 0$ section is shown), where the gain exists only if $|\omega| < \omega_{\max} = |48P/\beta_4|^{1/4}$ and the gain becomes maximum at two frequencies $\omega = \pm|24P/\beta_4|^{1/4}$. The MI gain spectrum reflects the exponential growth of the spectral-sideband perturbation, but it cannot characterize the spatiotemporal evolution after the growth reaches saturation, i.e., the nonlinear stage of MI.

## III. HETEROCLINIC STRUCTURE AND ITS TRANSITION

To describe the nonlinear stage of MI, the mode truncation approach is used by only considering the CW pump and the $\pm 1$ order sidebands at frequencies 0 and $\pm\omega$, respectively. We start by substituting in Eq. (1) the field:

$$A(z, t) = \sqrt{1-\eta} \exp(i\varphi_0) + \sqrt{\frac{\eta}{2}}[\exp(-i\omega t + i\varphi_1) + \exp(i\omega t + i\varphi_{-1})], \quad (2)$$

where $\eta(z) < 1$ is the total power of the first-order sidebands, and $\varphi_0(z)$ and $\varphi_{\pm 1}(z)$ are the pump and sideband phases, respectively. Since the total power of the field in Eq. (1) is equal to 1, we call $\eta$ and $1-\eta$ as the sideband fraction and the pump fraction, respectively. The frequency $\omega$ is henceforth limited to $0.5\omega_{\max} < \omega < \omega_{\max}$ to effectively prevent the emergence of higher-order sidebands. By substituting Eq. (2) into Eq. (1), we obtain the nonlinear wave evolution in the Hamiltonian form:

$$\dot\eta = -\frac{\partial H}{\partial \Delta\phi}, \quad \dot{\Delta\phi} = \frac{\partial H}{\partial \eta}, \quad (3a)$$

$$H(\Delta\phi, \eta) = (1-\eta)\eta\cos(2\Delta\phi) + \left(1 + \frac{\beta_4\omega^4}{24}\right)\eta - \frac{3\eta^2}{4}, \quad (3b)$$

where the dot stands for $d/dz$, $\Delta\phi = (\Delta\varphi_1 + \Delta\varphi_{-1} - 2\Delta\varphi_0)/2$ is the effective phase. The Hamiltonian $H$ can be depicted as contours on the ($\Delta\phi$, $\eta$) phase plane as shown in Figs. 1(c)–1(e), which give all possible dynamic trajectories of MI at a frequency $\omega$. It is obvious that the contours of $H$ have heteroclinic structure, and separatrixes exist [see the closed black and red curves in Figs. 1(c)–1(e)] dividing the phase plane into multiple regions. From the phase plane portraits of $H$, we can make a comprehensive analysis of the MI evolutions.

The Hamiltonian $H$ has four groups of stationary points ($\Delta\phi_e$, $\eta_e$) (solutions of $d\eta/dz = d\Delta\phi/dz = 0$) corresponding to four eigenmodes and being crucial for understanding the topology of heteroclinic structure for different $\omega$. Figure 1(b) shows the bifurcation diagram of such a stationary sideband fraction $\eta_e$ vs frequency $\omega$. We will introduce separately these stationary points below.

The first two groups of stationary points are ($\Delta\phi_e$, $\eta_e$) = [0 or $\pi$, $(4 + \beta_4\omega^4/12)/7$] and ($\Delta\phi_e$, $\eta_e$) = [$\pm\pi/2$, $|\beta_4|\omega^4/12$] corresponding to the maxima and minima of $H$, respectively. These two eigenmodes are stable because, when the Hamiltonian $H$ reaches its extremum, the contours of $H$ shrink into these two points ($\Delta\phi_e$, $\eta_e$). Figure 1(b) gives the relationships between $\eta_e$ and $\omega$ (see the black and red solid curves), and the two curves intersect when $\omega$ equals the critical frequency $\omega_{c1} = 8^{-1/4}\omega_{\max} \approx 1.565$. By comparing Figs. 1(c) and 1(d), it is noteworthy that, across $\omega_{c1}$, the topology of Hamiltonian $H$ changes suddenly, which has a profound impact on the spatiotemporal dynamics.

The third group of stationary points of ($\Delta\phi_e$, $\eta_e$) = [$\pm 0.5\cos^{-1}(\beta_4\omega^4/24 - 0.5)$, 1] indicating the unstable eigenmodes turn out to be the top vertexes of the contour $H(\Delta\phi, \eta) = \frac{1}{4} + \beta_4\omega^4/24$ in the frequency range of $\omega <$ $\omega_{c2} = \omega_{\max}/\sqrt{2} \approx 1.861$. It is worth noting that the contour of $H(\Delta\phi, \eta) = \frac{1}{4} + \beta_4\omega^4/24$ composes a heteroclinic separatrix [see the red line labeled separatrixes in Figs. 1(c) and 1(d)]. All the input conditions ($\Delta\phi_0$, $\eta_0$) only if locating on the separatrix can evolve asymptotically toward such an eigenmode, i.e., toward the state CW pump exhausts and only sidebands remain. It is very interesting that the heteroclinic separatrix here actually characterizes the frequency-conversion dynamics of a type of nonlinear dynamics with a modulated final state, which will be illustrated in Fig. 4(e).

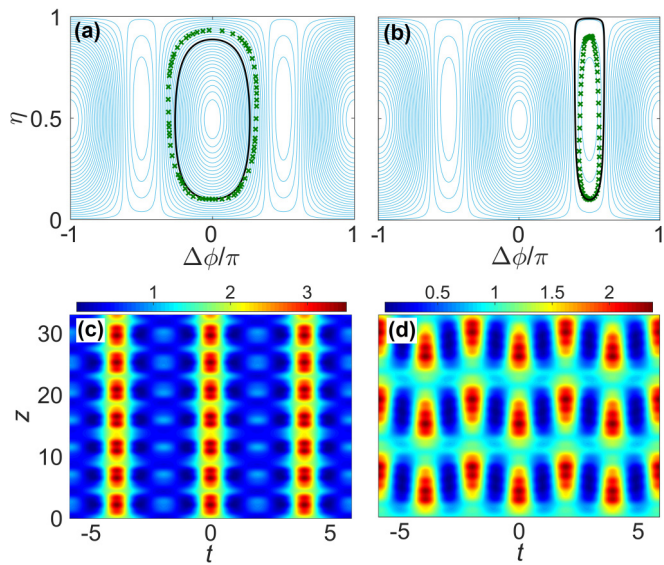

FIG. 2. Two types of unshifted Fermi-Pasta-Ulam (FPU) evolutions through numerical integration of Eq. (1) with initial effective phases: (c) $\Delta\phi_0 = 0$ and (d) $\Delta\phi_0 = 0.5\pi$. The corresponding projections of (a) Eq. (1) (discrete crosses) and (b) Eq. (3a) (black curve) on the phase plane ($\Delta\phi$, $\eta$). Here, $\eta_0 = 0.1$ and $\omega = 1.6$.

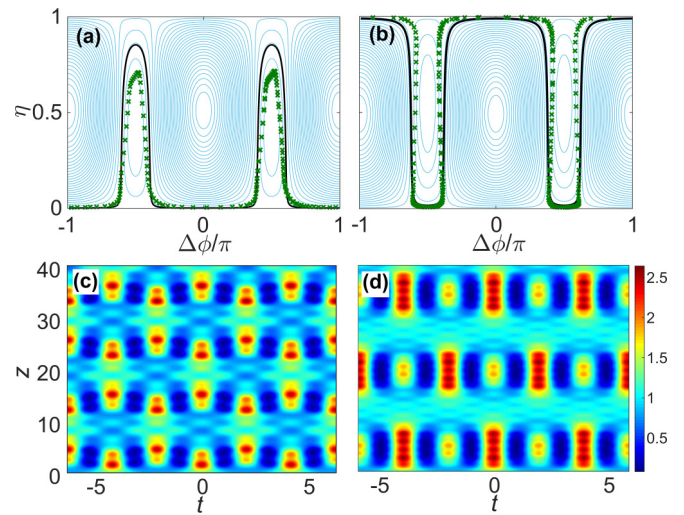

FIG. 3. Two types of staggered Fermi-Pasta-Ulam (FPU) evolutions from Eq. (1) numerical integration with modulational frequencies: (c) $\omega = 1.5$ and (d) $\omega = 1.6$. The corresponding projections of (a) Eq. (1) (discrete crosses) and (b) Eq. (3a) (black curve) on the $\Delta\phi$-$\eta$ plane. Here, $\eta_0 = 0.1$ and $\Delta\phi_0 = 0.39\pi$.

The fourth group of stationary points of ($\Delta\phi_e$, $\eta_e$) = [$\pm 0.5\cos^{-1}(\beta_4\omega^4/24 - 1)$, 0] correspond to the bottom vertexes of the heteroclinic separatrix of contour $H = 0$ [see the black lines in Figs. 1(c)–1(e)]. The stationary points here suggest also the unstable eigenmodes and only exist in the frequency range of $\omega < \omega_{\max}$, which agrees with the MI gain bandwidth from linear stability analysis. Being different from the third-group eigenmodes, when the input condition satisfies $H(\Delta\phi_0, \eta_0) = 0$, the nonlinear wave evolution pattern of Eq. (2) belongs to the ABs because only the CW-pump fraction remains in the asymptotical state. The corresponding ABs will be shown in Figs. 4(b)–4(d).

Three features of the heteroclinic structure are of crucial importance. First, when $\omega < \omega_{c2}$, the phase plane of $H$ has two different types of separatrixes individually possessing eigenmodes of $\eta_e = 0$ and 1 [see the black and red lines in Figs. 1(c) and 1(d)], which represent two types of wave evolutions with different final states. Second, across the first critical frequency $\omega_{c1}$, the two-type separatrixes suddenly switch their positions and proportions on the phase plane [compare the separatrixes in Figs. 1(c) and 1(d)]. This HST can bring substantial changes to both FPU recurrences and ABs patterns to be shown in Figs. 2–4. Third, when increasing $\omega$ ($\omega > \omega_{c1}$), the heteroclinic separatrix with stationary points $\eta_e = 1$ gradually shrinks, until the separatrix disappears at the second critical frequency $\omega = \omega_{c2}$, and then only the separatrix with stationary points $\eta_e = 0$ remains, and the heteroclinic structure [see Fig. 1(e)] becomes very similar to the well-known structure of the integrable NSE [4,22,23].

## IV. NONLINEAR DYNAMICS OF PATTERN FORMATION

The HST behavior of MI illustrated above has a profound influence on the long-term evolution of the nonlinear wave in Eq. (1). To show this, we numerically integrate Eq. (1) by taking the modulated plane wave as the initial condition: $A(0, t) = \sqrt{1 - \eta_0} + \sqrt{2\eta_0} \exp(i\theta_0) \cos(\omega t)$, where $\eta_0$ is the initial sideband fraction, and $\theta_0$ is taken as the overall initial effective phase $\theta_0 = \Delta\phi_0 = \Delta\phi(0)$. On the other hand, we also numerically integrate Eq. (3a) by taking the same values of $\eta_0$ and $\theta_0$ for comparison. In the following discussion, we mainly focus on the optical-wave dynamics in the frequency range of $\omega < \omega_{c2}$ because the dynamic behavior for $\omega > \omega_{c2}$ have no substantial difference from the case of integrable NSE.

### A. Fermi-Pasta-Ulam recurrences

We first perform the nonlinear-wave evolutions with dynamic trajectories inside the heteroclinic separatrixes for $\omega < \omega_{c2}$, which exhibit two types of unshifted FPU recurrences, as shown in Fig. 2. Figures 2(c) and 2(d) display their dynamics with the initial phases of $\Delta\phi_0 = 0$ and $0.5\pi$, respectively. Their dynamic trajectories are depicted in Figs. 2(a) and 2(b), where the discrete crosses and black curve represent the numerical results from Eqs. (1) and (3a), respectively. Obviously, the numerical results by Eq. (3a) have distinct deviations from those by Eq. (1), which is induced by the neglected higher-order sidebands [same reason for the deviations in Figs. 3(a) and 3(b)]. The two types of unshifted FPU recurrences have dynamic trajectories being located inside different heteroclinic separatrixes in the phase plane. Note that we value the frequency as $\omega = 1.6$ in the frequency range $\omega_{c1} < \omega < \omega_{c2}$ in Fig. 2, whereas the very similar FPU-recurrence patterns also occur for $\omega < \omega_{c1}$.

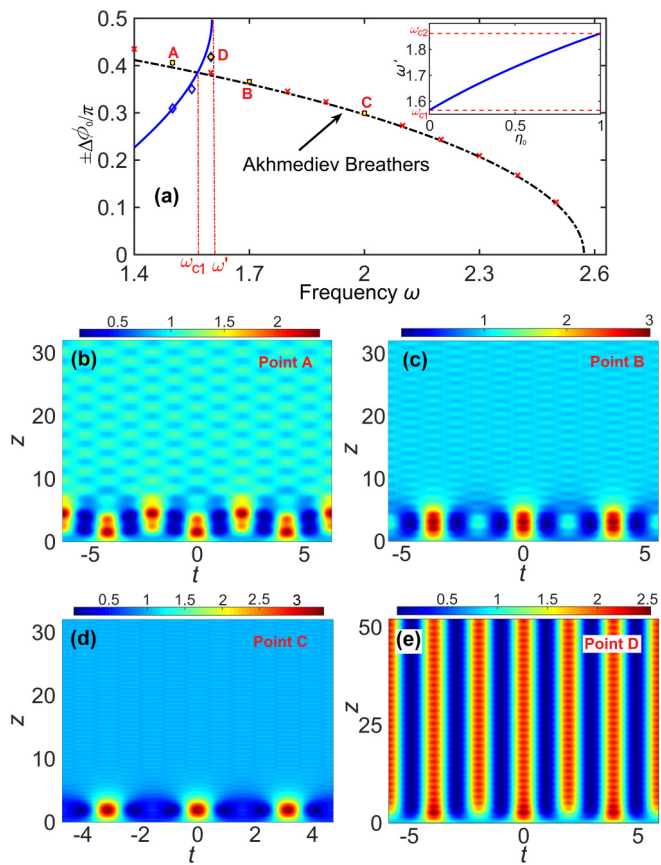

FIG. 4. (a) The initial phase $\Delta\phi_0$ vs frequency $\omega$ for the nonlinear-wave evolutions on the heteroclinic separatrixes with $\eta_e = 1$ (solid-blue line) and on the separatrixes with $\eta_e = 0$ (dash-dotted black line), compared with the ones by numerical integration of Eq. (1) (discrete crosses and rhombuses). Inset: $\omega'$ vs input sideband fraction $\eta_0$ (the horizontal dashed lines stand for the critical frequencies of $\omega_{c1}$ and $\omega_{c2}$). The corresponding numerical evolutions of A–D points in (a) at frequencies (b) $\omega = 1.5$, (c) $\omega = 1.7$, (d) $\omega = 2$, and (e) $\omega = 1.6$. They have initial phases of $\Delta\phi_0 = 0.406328\pi$, $0.366239\pi$, $0.299573\pi$, and $0.418065\pi$, respectively. All of the above have $\eta_0 = 0.1$.

Next, we discuss the dynamic orbits outside the heteroclinic separatrix in the phase plane, which correspond to two types of staggered FPU recurrences with phase spanning continuously the full range $(-\pi, \pi)$, as shown in Fig. 3. All the numerical simulations in Fig. 3 have the same initial values of $\eta_0 = 0.1$ and $\Delta\phi_0 = 0.39\pi$ except different frequencies of $\omega = 1.5$ for Figs. 3(a) and 3(c) and $\omega = 1.6$ for Figs. 3(b) and 3(d). Thus, the two types of staggered FPU recurrences are separately arisen in frequency ranges of $\omega < \omega_{c1}$ [see Figs. 3(a) and 3(c)] and $\omega_{c1} < \omega < \omega_{c2}$ [see Figs. 3(b) and 3(d)]. According to their pattern type, the recurrences displayed in Figs. 3(c) and 2(d) should belong to one same type of FPU recurrence, and we name them type-A FPU, while the recurrences displayed in Fig. 3(d) and 2(c) belong to another type of FPU, and we name them type-B FPU. Therefore, one influence of HST across $\omega = \omega_{c1}$ on nonlinear wave evolution is the staggered-FPU switching from type A to B.

### B. Nonlinear waves on the heteroclinic separatrixes

Now we focus on the nonlinear wave excitations with trajectories on the heteroclinic separatrixes. To illustrate their excitation conditions clearly, the curves of initial phase $\Delta\phi_0$ vs $\omega$ are drawn in Fig. 4(a) for fixed $\eta_0 = 0.1$. The nonlinear waves that have orbits along the heteroclinic separatrixes with $\eta_e = 0$ [i.e., the black lines in Figs. 1(c)–1(e)] and along the separatrixes with $\eta_e = 1$ [i.e., the red lines in Figs. 1(c) and 1(d)] are generated under the initial phases that satisfy equations $H(\Delta\phi_0, \eta_0) = 0$ and $\frac{1}{4} + \beta_4\omega^4/24$, respectively. Their solutions of the equations are depicted as dash-dotted and solid blue lines in Fig. 4(a) and agree well with the numerical results. The numerical simulation results show that the separatrixes with $\eta_e = 0$ correspond to the AB evolutions with CW final states despite the higher-order sideband-induced weak modulations as shown in Figs. 4(b)–4(d). According to their dynamic patterns, these ABs can be classified as type-A [see Fig. 4(b)] and type-B [see Figs. 4(c) and 4(d)] ABs in the frequency ranges of $\omega < \omega_{c1}$ and $\omega > \omega_{c1}$, respectively. Note that the definitions of type-A and type-B ABs are like the ones of FPU recurrences above. We make both patterns of Figs. 4(c) and 4(d) belong to type-B ABs because the AB like Fig. 4(c) can gradually evolve into the one like Fig. 4(d) with the increase of $\omega$. The orbits of type-A and type-B ABs in the phase plane actually run along the separatrixes like the black lines in Figs. 1(c) and 1(d), respectively. Thus, there is a sudden switch from type-A to type-B ABs across $\omega = \omega_{c1}$ [also compare the black line labeled separatrixes in Figs. 1(c) and 1(d)].

It is very important that the separatrixes with $\eta_e = 1$ represent another type of nonlinear wave as shown in Fig. 4(e), which shows a modulated final state. The numerical simulations show that, although such waves have dynamic trajectories along the analogous separatrixes of red lines in Figs. 1(c) or 1(d), they always have the evolution dynamics like Fig. 4(e). Note that these waves can only exist in the frequency range of $\omega < \omega'$, where $\omega'$ is $\eta_0$ dependent and is solved as $\omega' = \omega_{c1}(1 + \eta_0)^{1/4}$ by the equation $H(\pi/2, \eta_0) = \frac{1}{4} + \beta_4\omega^4/24$. When the initial sideband fraction $\eta_0$ increases from 0 to 1, the upper-edge frequency $\omega'$ is increased from $\omega_{c1}$ to $\omega_{c2}$ as shown in the inset in Fig. 4(a). That is because the separatrix with $\eta_e = 1$ in the frequency range $\omega_{c1} < \omega < \omega_{c2}$ gradually shrinks with the increase of $\omega$, until which disappears at $\omega = \omega_{c2}$. Moreover, regardless of the value $\eta_0$, the two curves in Fig. 4(a) always intersect at $\omega = \omega_{c1}$ because the two types of heteroclinic separatrixes always switch their positions and proportions at $\omega_{c1}$.

It is necessary to further compare the ABs and the wave along separatrixes with $\eta_e = 1$ from the perspective of frequency conversion. Figure 5 shows their frequency-conversion dynamics by the numerical integration of Eq. (1). Both types of ABs experience a complex power exchange between the pump and sidebands until the pumps dominate the propagation after $z = 10$, as shown in Figs. 5(a) and 5(b). Thus, the final states of ABs are plane waves with spectra as shown in the insets of Figs. 5(a) and 5(b). However, the wave along the separatrixes with $\eta_e = 1$ has a completely different frequency conversion that the pump is almost completely converted to the sidebands after $z = 6$, as shown in Fig. 5(c). Therefore, the evolution has a final state that only sidebands remain in the spectrum [see the inset of Fig. 5(c)], which presents a cosine distribution in time domain as shown in Fig. 4(e).

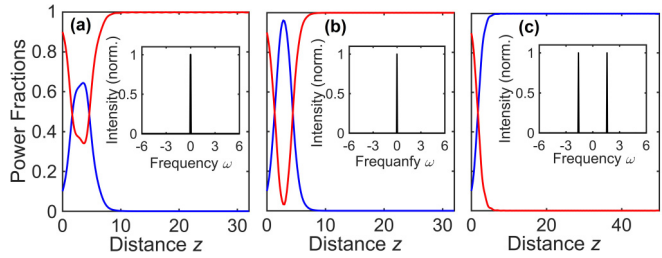

FIG. 5. The evolutions of extracted pump (red line) and sideband power (blue line) fractions corresponding to (a) the type-A AB in Fig. 4(b), (b) the type-B AB in Fig. 4(c), and (c) the dynamics in Fig. 4(e). Insets: Their final-state spectra.

## V. CONCLUSIONS

In conclusion, we have revealed the HST behaviors of pure-quartic MI at two critical frequencies. This guided us to explore the rich nonlinear dynamics in the biharmonic NSE, especially the discovery of two types of nonlinear waves: the type-A FPU and AB patterns, and the waves along the heteroclinic separatrixes with $\eta_e = 1$. From our work, we hope to stimulate further theoretical and experimental studies intending to broaden and deepen the understanding of MI.

This paper was supported by the National Natural Science Foundation of China (Grants No. 12004309, No. 12175178, No. 11875220, and No. 12047502), and by Scientific Research Program Funded by Shaanxi Provincial Education Department (Grant No. 20JK0947), and the Major Basic Research Program of Natural Science of Shaanxi Province, (Grant No. 2017KCT-12).